\documentclass[amssymb,prb,twocolumn,showpacs,superscriptaddress]{revtex4}
\usepackage{amsmath}
\usepackage{graphicx}
\usepackage{verbatim}
\usepackage{graphics}
\usepackage{units}
\usepackage{color}
\usepackage{ulem}
\newcommand{\bea}{\begin{eqnarray}}
\newcommand{\eea}{\end{eqnarray}}
\newcommand{\be}{\begin{equation}}
\newcommand{\ee}{\end{equation}}

\begin{document}

\title{
Superexchange blockade in triple quantum dots
}
\author{Rafael S\'anchez}
\affiliation{Instituto de Ciencia de Materiales de Madrid, CSIC, Cantoblanco, 28049, Madrid, Spain}
\author{Fernando Gallego-Marcos}
\affiliation{Instituto de Ciencia de Materiales de Madrid, CSIC, Cantoblanco, 28049, Madrid, Spain}
\affiliation{Institut f\"ur Theoretische Physik, Technische Universit\"at Berlin, D-10623 Berlin, Germany}
\author{Gloria Platero}
\affiliation{Instituto de Ciencia de Materiales de Madrid, CSIC, Cantoblanco, 28049, Madrid, Spain}
\begin{abstract}
We propose  the interaction of two electrons in a triple quantum dot as a minimal system to control long range superexchange transitions. These are probed by transport spectroscopy. Narrow resonances appear indicating the transfer of charge from one side of the sample to the other with the central one being occupied only virtually. We predict that two different intermediate states establish the two arms of a one dimensional interferometer. We find configurations where destructive interference of the two superexchage trajectories totally blocks the current through the system. We emphasize the role of spin correlations giving rise to lifetime-enhanced resonances.
\end{abstract}
\maketitle

{\it Introduction.} Transitions mediated by long range quantum coherence in two or more particle systems are an essential concept in many different fields.  Superexchange, the interaction of orbitals whose overlap is small but is mediated by intermediate virtual states, was introduced by Pauling in his resonance theory of the molecular bond~\cite{pauling}. Delocalization due to long range electron transfer mechanisms is responsible for donor-acceptor reactions through bridge states~\cite{gray,ratner} relevant for molecules as complex as photosynthetic centers~\cite{hu} or DNA~\cite{jortner,giese}.
In the solid state, seminal works by Zener~\cite{zener} and Anderson~\cite{anderson1959} introduced long range exchange interactions to explain transport and order in magnetic compounds. Related ideas led to models of the Kondo problem~\cite{anderson1961,kondo}. Resonance valence bond models~\cite{andersonCu} have found recently an increased interest in the context of topological phases in triangular lattices~\cite{moessner}.



The complex physics involved in the above mentioned systems can be unraveled by investigating simpler configurations that can be realized experimentally. For that purpose, quantum dot arrays are ideal for their scalability, high degree of tunability and long coherence times~\cite{barthelemy}. Coupled quantum dots behave as artificial molecules and their coupling is naturally described by hopping Hamiltonians. These characteristics nominate them for simulations of chemical reactions~\cite{smirnov} or lattice models~\cite{manousakis,byrnes,seo}. The interplay of charge and spin correlations introduces unique transport dynamics as the mesoscopic Kondo effect~\cite{gordon} or Pauli spin blockade~\cite{ono}. The impressive gate control of few electron triple quantum dots~\cite{gaudreau,schroer,ghislain,amaha-spin} has succeeded the operation of three electron exchange-only qubits~\cite{gaudreau-qubit,studenikin,laird}. In situations where tunneling to the centre dot is energetically forbidden, superexchange is responsible for the indirect coupling of the two outer quantum dots, mediated by virtual transitions through the middle one. Evidences of such transitions have been recently reported in the form of sharp current resonances~\cite{maria,qp12} and by real-time charge detection~\cite{braakman}. Thus quantum dots offer not only a way to experimentally control superexchange but also the possibility to explore new phenomena based on long range interactions~\cite{reilly,saraga}.

\begin{figure}[b]
\begin{center}
\includegraphics[width=\linewidth,clip] {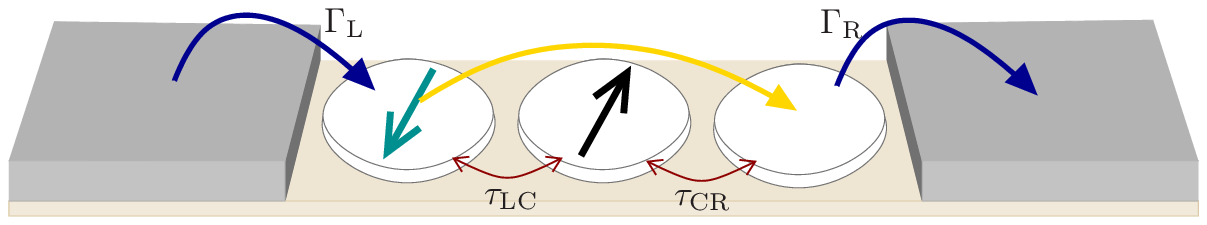}\\
\vspace{0.3cm}
\includegraphics[width=0.9\linewidth,clip] {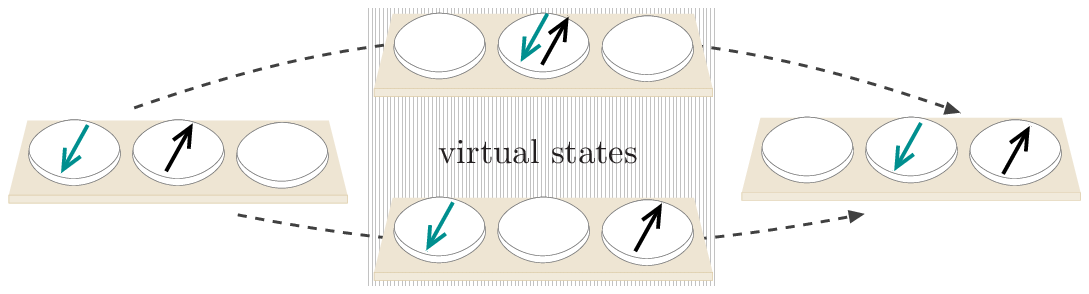}
\end{center}
\caption{\label{esquema} A triple quantum dot with superexchange mediated transport. Electrons tunnel from the source lead into the left dot and from the right dot into the drain lead with rates $\Gamma_\text{L,R}$. Interdot tunneling is described by the hopping terms $\tau_\text{LC}$ and $\tau_\text{CR}$. When states (1,1,0) and (0,1,1) are degenerate, charge is transferred via higher-order tunneling processes in which the intermediate states are only virtually occupied. The two intermediate states, whose energy is tunable by gate voltages, define the two arms of a superexchange interferometer.
}
\end{figure}

Here we investigate the minimal system with long range superexchange interactions affected by charge and spin correlations. It requires three sites and two electrons. In particular, two-particle correlations introduces a mechanism for the quantum interference of superexchange transitions. 
At the degeneracy of $(N_\text{L},N_\text{C},N_\text{R})$ = (1,1,0) and (0,1,1) states ---$N_l$ being the number of electrons in quantum dot $l$--- charge is delocalized between the two edge dots via the virtual occupation of {\it two} possible intermediate states, (0,2,0) and (1,0,1), which are detuned, as sketched in Fig.~\ref{esquema}. 

We focus on a configuration where the two different virtual transitions coexist and lead to interference. Remarkably, we find conditions where the destructive interference of transitions through the (1,0,1) and (0,2,0) branches completely cancels the transport, what we term superexchange blockade. We emphasize the role of spin correlations. The two-path interference only occurs for singlet states: Pauli exclusion principle avoids triplets to tunnel into the (0,2,0) state. As a consequence, at the condition for superexchange blockade, triplets contribute to transport assisted by long range tunneling through (1,0,1), while the occupation of singlet states cancels the current. This mechanism is in utter contrast with spin blockade in coupled quantum dots, where triplets block the current.

\begin{figure}[t]
\begin{center}
\includegraphics[width=0.9\linewidth,clip] {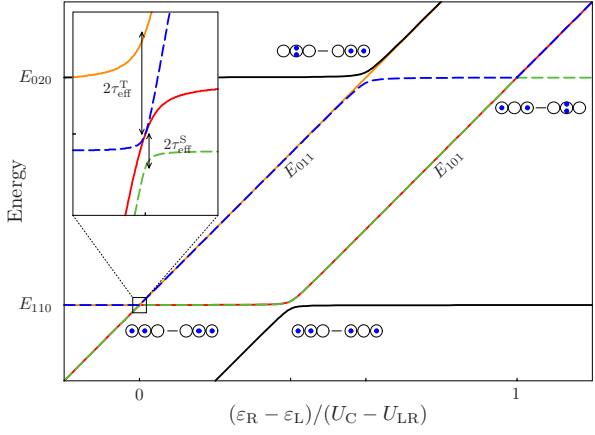}
\end{center}
\caption{\label{eigen} Eigenenergies of the two electron sector of $\hat{H}_{\text{TQD}}$. Anticrossings develop at the degeneracies of states with different charge distribution. In our configuration, where (1,0,1) and (0,2,0) do not carry current, transport will occur only around the (1,1,0)--(0,1,1) crossing. As shown in the inset, spin correlations introduce different couplings $\tau_\text{eff}^\text{S}$ and $\tau_\text{eff}^\text{T}$ for singlet (dashed) and triplet states (solid lines). 
Parameters (in meV): $\tau_\text{LC}{=}\tau_\text{CR}{=}0.005$, $U_\text{R}{=}0.8$, $U_\text{LC}{=}U_\text{CR}{=}0.5$, $U_\text{LR}{=}0.3$.
}
\end{figure}

{\it Model.} We describe the triple quantum dot with the three-site Anderson Hamiltonian ($i{=}\text{L,C,R}$):
\bea
\label{ham}
\hat{H}_{\text{TQD}} &= &\sum_{i\sigma}\varepsilon_{i}\hat{c}^{\dagger}_{i\sigma}
\hat{c}_{i\sigma}+\sum_{i}U_{i}\hat{n}_{i\uparrow}\hat{n}_{i\downarrow}
+\frac{1}{2}\sum_{i\neq j}U_{ij}\hat{n}_{i}\hat{n}_{j}\nonumber\\
&&-\sum_{i\neq j,\sigma}\tau_{ij}\hat{c}^{\dagger}_{i\sigma}\hat{c}_{j\sigma}+\text{H.c.},
\eea
with on-site energy levels $\varepsilon_{i}$, intra- and inter-dot Coulomb repulsion $U_i$ and $U_{ij}$, and nearest neightbour hopping $\tau_{ij}$. The outer dots are tunnel-coupled to fermionic reservoirs $\hat{H}_{\text{leads}} {=} \sum_{l,k\sigma}\varepsilon_{lk}\hat{d}^{\dagger}_{lk\sigma}\hat{d}_{lk\sigma}$ by $\hat{H}_{\text{tun}} = \sum_{l,k\sigma}\gamma_{l}\hat{d}^{\dagger}_{lk\sigma}
\hat{c}_{l\sigma}+\text{H.c.}$ We consider a large Coulomb interaction with up to two electrons in the system. Our system is hence described by five charge distribution states: $|0,\sigma,0\rangle=\hat{c}_{C\sigma}^\dagger|0\rangle$, $|\sigma,\sigma',0\rangle=\hat{c}_{\text{L}\sigma}^\dagger\hat{c}_{\text{C}\sigma'}^\dagger|0\rangle$, $|\sigma,0,\sigma'\rangle=\hat{c}_{\text{L}\sigma}^\dagger\hat{c}_{\text{R}\sigma'}^\dagger|0\rangle$, $|0,\sigma,\sigma'\rangle=\hat{c}_{\text{C}\sigma}^\dagger\hat{c}_{\text{R}\sigma'}^\dagger|0\rangle$ and $|0,2,0\rangle=\hat{c}_{\text{C}\uparrow}^\dagger\hat{c}_{\text{C}\downarrow}^\dagger|0\rangle$. If a high bias voltage, $(\mu_\text{L}-\mu_\text{R})/e$, is applied between the left and right terminals, with $\mu_l$ the chemical potential of lead $l$, current is unidirectional from left to right. In our configuration, with $\mu_\text{L(R)}>\varepsilon_\text{L(R)}+U_\text{LR}$, (1,0,1) and (0,2,0) states do not carry current. Tunneling through the leads thus occurs via the transitions $|0,\sigma,0\rangle\rightarrow|\sigma',\sigma,0\rangle$, and  $|0,\sigma,\sigma'\rangle\rightarrow|0,\sigma,0\rangle$, with $|\sigma,0,\sigma'\rangle$ and $|0,{\uparrow}{\downarrow},0\rangle$ acting as the intermediate states for the charge transfer within the triple quantum dot. Clearly, the current through the system will be enhanced close to the resonances of states with charge distributions differing in the position of one electron. This is also the case for the crossing of $|\sigma',\sigma,0\rangle$ and $|0,\sigma,\sigma'\rangle$ states, i.e. when $E_{110}\approx E_{011}$, even if the intermediate states are far in energy, see Fig.~\ref{eigen}. Then, charge delocalization requires higher-order transitions. That is the superexchange-mediated transport that we are interested in. 

\begin{figure}[t]
\begin{center}
\includegraphics[width=0.9\linewidth,clip] {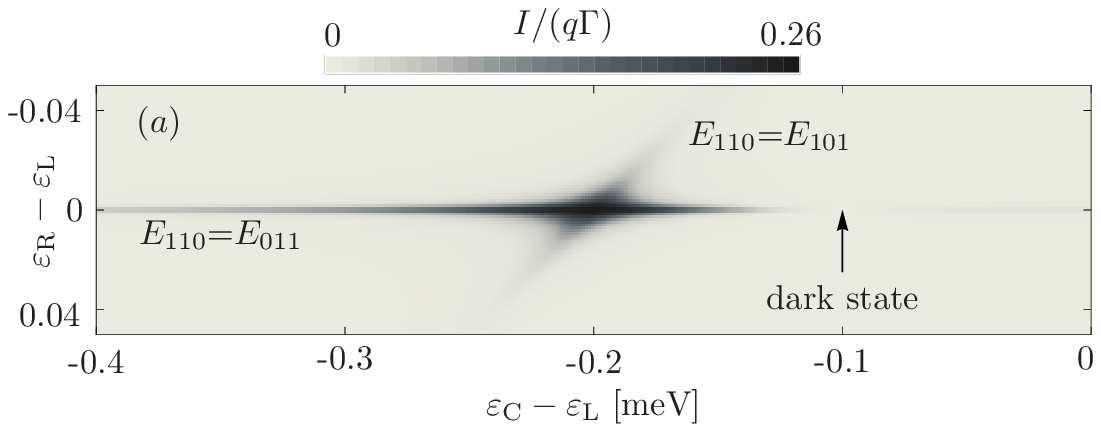}
\includegraphics[width=0.9\linewidth,clip] {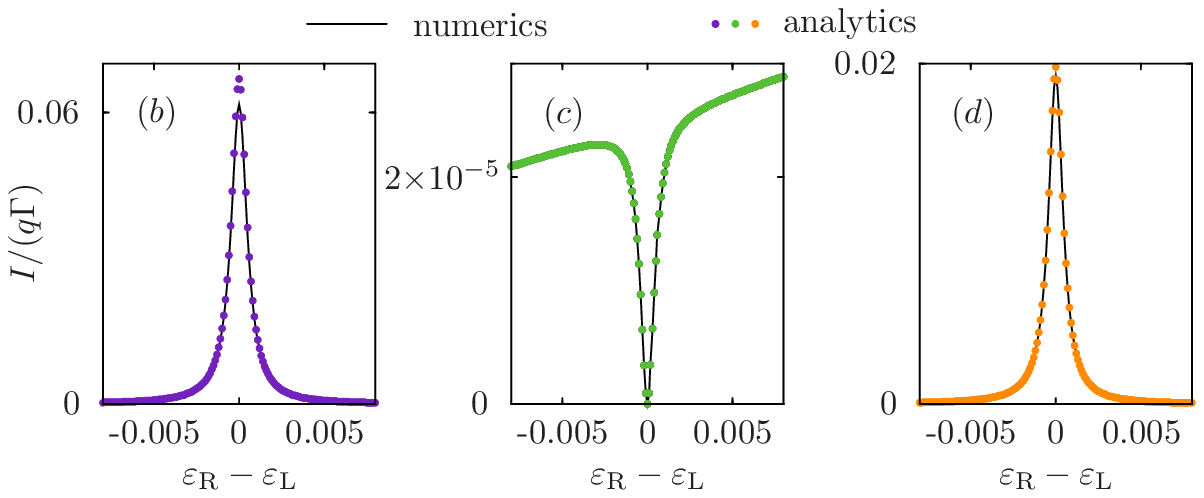}
\end{center}
\caption{\label{curr} Current as a function of the left-center and left-right detunings for the same parameters of Fig.~\ref{eigen}. (a) A large resonance appears at the triple crossing (1,1,0)--(1,0,1)--(0,1,1). A narrow resonance at the degeneracy (1,1,0)--(0,1,1) survives far from the crossing maintained by superexchange transitions. At the condition $E_{110}{-}E_{020}{=}{-}2(E_{110}{-}E_{101})$ a dark state is formed that cancels the current. Here $\Gamma_\text{L}=\Gamma_R=\unit[1]{\mu eV}$. A good agreement is found between the full numerical calculation and the analytical expression for the L-R resonance, $I_\text{LXR}$ (see text), for (b) $\varepsilon_\text{C}-\varepsilon_\text{L}=-0.4$, (c) $-0.1$ ---showing the superexchange current blockade due to the formation of a dark state, Eq.~\eqref{darkst}--- and (d) \unit[0]{meV}.
}
\end{figure}

The charge current thus serves to probe the superexchange transitions in the system. We calculate it from the stationary solution of the quantum master equation for the reduced density matrix of the quantum dot system~\cite{mariaSB}: $\dot{\hat\rho}=-i\hbar^{-1}[\hat{H}_{\text{TQD}},\hat\rho]+{\cal L}_\Gamma\hat\rho=0$. The Liouvillian superoperator ${\cal L}_\Gamma$ includes the tunneling events through the leads. In the weak coupling regime, where we can neglect cotunneling contributions, they are described by Fermi's golden rule, $\Gamma_l=\frac{2\pi}{\hbar}|\gamma_l|^2\nu_l$, with $\nu_l$ being the density of states in lead $l$. The current to the right lead is given by the occupation probability of (0,1,1) states: $I=q\Gamma_\text{R}\sum_{\sigma\sigma'}\langle 0,\sigma,\sigma'|\hat\rho|0,\sigma,\sigma'\rangle$. The result is shown in Fig.~\ref{curr}, where, as expected, a large peak appears when the sequence $|\sigma,\sigma',0\rangle{\rightarrow}|\sigma,0,\sigma'\rangle{\rightarrow}|0,\sigma,\sigma'\rangle$, carrying charge from the left to the right dot, occurs resonantly. On the other hand, no feature is observed at the resonance of states $|\sigma,\sigma',0\rangle$, $|0,2,0\rangle$ and $|0,\sigma,\sigma'\rangle$: the occupation of a triplet state prevents this transition (spin blockade).
However, the current is not totally canceled as virtual tunneling through (1,0,1) bypasses the blockade.

Most interestingly, a narrow resonance appears along the condition $E_{110}=E_{011}$ (for $\varepsilon_\text{L}-\varepsilon_\text{R}=U_\text{CR}-U_\text{LC}$) due to superexchange interactions. As we show below, it involves charge being transferred from the left to the right dot without ever changing the occupation of the center dot~\cite{maria,qp12}. Importantly, the resonance is cancelled at a particular configuration where, as we discuss below, destructive interference leads to a dark state, cf. Figs.~\ref{curr}(a) and (c). This is the evidence of superexchange blockade.
 
{\it Left-right resonance.} In order to better understand the dynamics at this resonance, we perform a perturbative expansion of the two-electron sector of $\hat{H}_{\text{TQD}}$ around $E_{110}\approx E_{011}$ for the case where the detunings $\delta_{101}=E_{110}-E_{101}$ and $\delta_{020}=E_{110}-E_{020}$ are large compared to the interdot hopping, $\tau_{ij}\ll|\delta_{101}|,|\delta_{020}|$. Introducing the operators ${\cal P}=\sum_{\sigma,\sigma'}(|{\sigma},{\sigma'},0\rangle\langle{\sigma},{\sigma'},0|+|0,{\sigma},{\sigma'}\rangle\langle0,{\sigma},{\sigma'}|)$ and ${\cal Q}=1-{\cal P}$, we project out the intermediate states and obtain the effective Hamiltonian $\hat{H}_\text{eff}={\cal P}\hat{H}_{\text{TQD}}{\cal P}+{\cal P}\hat{H}_{\text{TQD}}{\cal Q}[E-{\cal Q}\hat{H}_{\text{TQD}}{\cal Q}]^{-1}{\cal Q}\hat{H}_{\text{TQD}}{\cal P}$. It is convenient for the following to introduce the singlet and triplet superpositions $\Lambda_{q}$=$\{S_{q},T_{q}^\alpha\}$, with $q$=\{LC,CR\} and $\alpha$=$\{0,\pm\}$. 
Thus, we get $\hat{H}_\text{eff}=\sum_qE_{\Lambda_q}|\Lambda_{q}\rangle\langle\Lambda_{q}|+\hat{H}_\text{LXR}$, 
with $E_{T_q}=E_{110}+\tau_{\bar q}^2/\delta_{101}$ and $E_{S_q}=E_{T_q}+2\tau_{q}^2/\delta_{020}$. Note that the energy is shifted due to two different second order processes: one affects both the singlet and triplet states and is induced by charge fluctuations through barrier $\bar q\ne q$; the other one is due to (direct) Heisenberg exchange through $q$ and is responsible for singlet-triplet splitting, see the inset in Fig.~\ref{eigen}. 
Most importantly, 
\be
\label{hlxr}
\hat{H}_\text{LXR}=\tau_\text{eff}^\text{S}|S_\text{LC}\rangle\langle S_\text{CR}|+\tau_\text{eff}^\text{T}|T_\text{LC}^\alpha\rangle\langle T_\text{CR}^\alpha|+\text{H.c.}
\ee
describes transitions where charge is delocalized between the left and right dots via the second order tunneling couplings $\tau_\text{eff}^\text{S}=\tau_\text{LC}\tau_\text{CR}(\delta_{101}^{-1}+2\delta_{020}^{-1})$ and $\tau_\text{eff}^\text{T}=\tau_\text{LC}\tau_\text{CR}\delta_{101}^{-1}$. These transitions are responsible for the narrow resonance at the crossing of (1,1,0) and (0,1,1) states.
Note that, for involving different transitions, the effective coupling is not the same for singlet and triplet states. As they depend on energy detuning, they can be controlled by the gate voltages applied to the quantum dots. The anticrossing has a different gap and curvature in each case, see inset of Fig.~\ref{eigen}, what has been invoked as a signature of hybrid states in recent experiments~\cite{amaha}.

The system is then effectively reduced to 10 states (including those with a single particle), so the master equation can still be solved analytically. We calculate the current which has a Lorentzian shape, $I_\text{LXR}=I_0W^2[(E_{110}-E_{011})^2+W^2]^{-1}$, of height $I_0=q4\Gamma W^{-2}(\tau_\text{eff}^\text{S}\tau_\text{eff}^\text{T})^2/(3\tau_\text{eff}^\text{S}{}^2+\tau_\text{eff}^\text{T}{}^2)$ and width:
\be
W=\sqrt{\frac{\Gamma^2}{4}+\frac{10(\tau_\text{eff}^\text{S}\tau_\text{eff}^\text{T})^2}{3\tau_\text{eff}^\text{S}{}^2+\tau_\text{eff}^\text{T}{}^2}}.
\ee 
As shown in Fig.~\ref{curr}, it agrees well with the full calculation using $\hat{H}_\text{TQD}$, Eq.\eqref{ham}. 

We can now explore some interesting configurations analytically:
(i) at $\delta_{101}=-\delta_{020}$, the singlet and triplet states have the same effective coupling: $\tau_\text{eff}^\text{S}=\tau_\text{eff}^\text{T}$. One thus easily verifies that current flows through the eigenstate superpositions $|\Lambda_\text{LC}\rangle\pm|\Lambda_\text{CR}\rangle$, for either singlets or triplets. Note furthermore that they involve the transport of maximally entangled two-electron superpositions along the quantum dot chain: Any Bell superposition of two electrons in one edge and in the centre dot, $\Psi_q^i=\{S_{q},T_{q}^0,T_{q}^+\pm T_{q}^-\}$, can be transferred to the other pair of dots, e.g. $|\Psi_\text{LC}\rangle\rightarrow|\Psi_\text{CR}\rangle$, by the appropriate application of voltage pulses. This makes our mechanism a candidate for the coherent transfer by adiabatic passage (CTAP) of Bell states, of importance for quantum information processing~\cite{fabian,jan}.  (ii)  Most interestingly for us here, at $\delta_{101}=-\delta_{020}/2$, the two singlet virtual paths interfere destructively and $\tau_\text{eff}^\text{S}=0$. This is not the case for triplet trajectories: they can only be transmitted along the (1,0,1) branch and hence do not interfere. 
At this point, the singlet $|S_\text{LC}\rangle$ does not contribute to transport and will block the current, as we further discuss below. We remark that the system behaves as a one-dimensional two-particle interferometer. The destructive interference of superexchange trajectories is unique to our setup for involving discrete states only. Note the difference with virtual (cotunneling) transitions involving the leads~\cite{averin}: the contributions of many trajectories due to the wide density of states distribution sum to give a finite rate.

{\it Superexchange blockade.} The effect of the destructive interference of singlet trajectories is evident in the transport through the system at the condition $\delta_{020}=-2\delta_{101}=\Delta$. 
Whenever the electrons in the centre and source dots form a spin singlet, current will be suppressed, see Figs.~\ref{curr}(a) and \ref{curr}(c). 
By diagonalizing the Hamiltonian \eqref{ham} of the total system at this point, and considering for simplicity $\tau_\text{LC}=\tau_\text{CR}=\tau$, we find the eigenstate 
\be
\label{darkst}
|\text{DS}\rangle=\Delta|S_\text{LC}\rangle-\tau|0,{\uparrow}{\downarrow},0\rangle+2^{-1/2}\tau|S_\text{LR}\rangle
\ee
which contains no contribution of $|S_\text{CR}\rangle$ and therefore does not carry current. Hence, the system evolves toward a stationary pure state $\rho_\text{st}=|\text{DS}\rangle\langle\text{DS}|$.
The dark state \eqref{darkst} is formed via the destructive interference of two-particle trajectories in one dimension. Quite differently, transport dark states proposed (and not measured) so far require more complicated setups: either the spatial separation of single-electron trajectories in two dimensional arrays~\cite{alfredo,cpt} or excited states in ac driven configurations~\cite{brandes,darkbell}. Our prediction of the vanishing current at the long-range resonance is within present experimental reach~\cite{braakman} and paves the way to the detection of a transport dark state.
We want to stress that the cancellation of the current is exact to all orders in the hopping, not restricted to the perturbative expansion~\eqref{hlxr}.

The superexchange blockade induced by the formation of the dark state $|\text{DS}\rangle$ is opposite to the spin blockade effect, where it is the occupation of triplet states that blocks the current~\cite{ono,maria,qp12}. In our case, it is indeed the spin blockade mechanism which prevents triplets to block the current by eliminating one of the interference paths.  


\begin{figure}[t]
\begin{center}
\includegraphics[width=0.9\linewidth,clip] {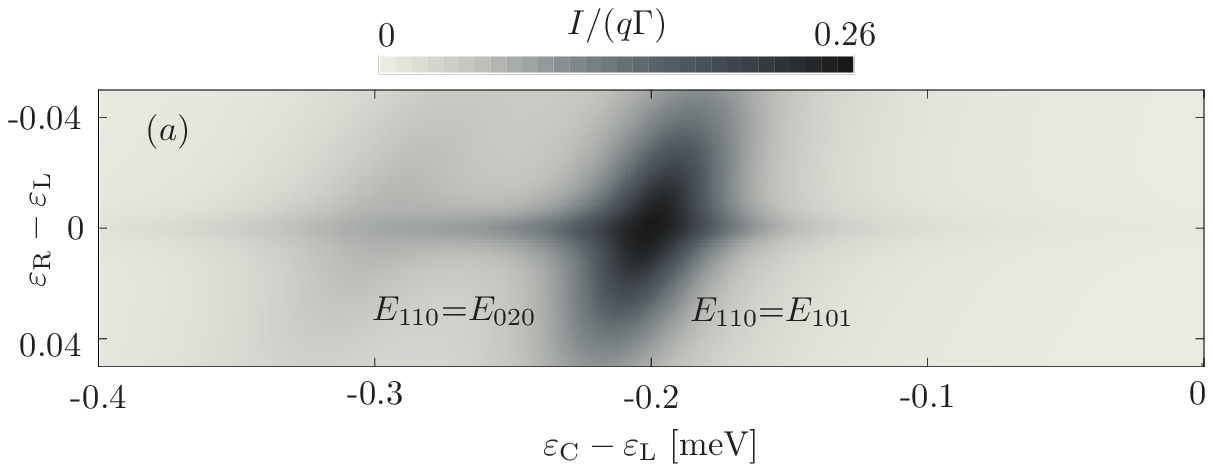}
\includegraphics[width=0.9\linewidth,clip] {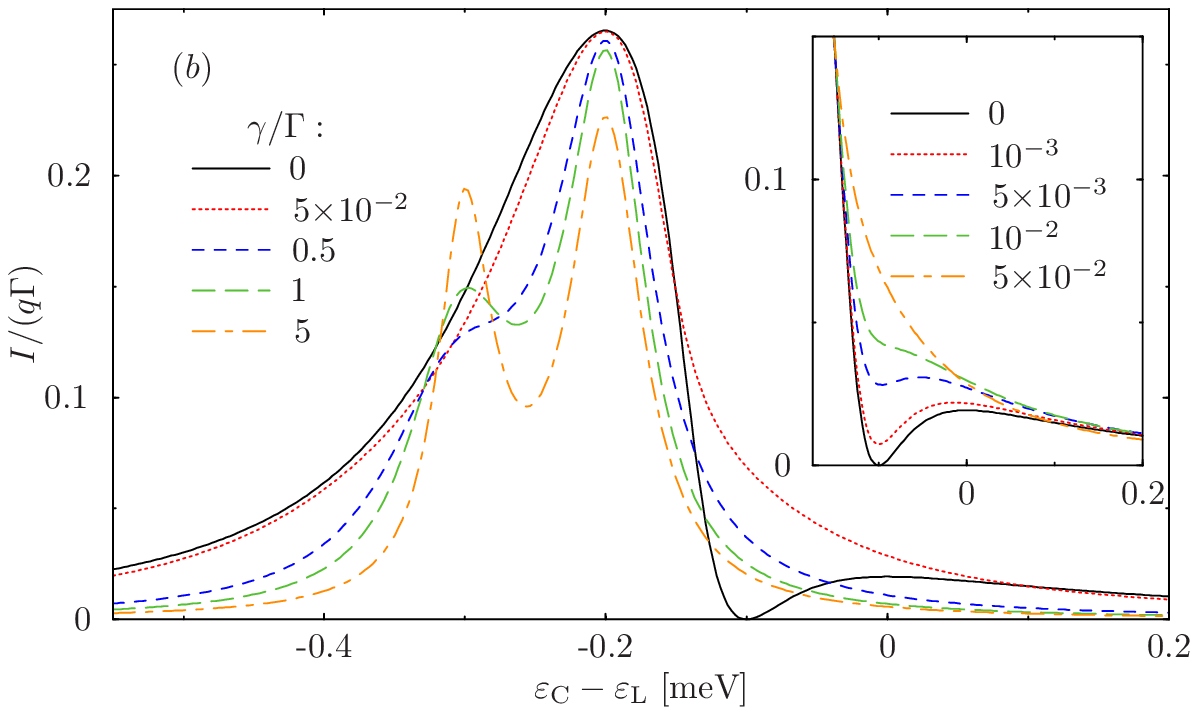}
\end{center}
\caption{\label{curr-rel} (a) Current as a function of the left-centre and left-right detunings in the presence of finite spin decoherence. Spin relaxation and decoherence rates are $T_1^{-1}=5{\times}10^{-3}\gamma$ and $T_2^{-1}=\gamma$, respectively. (b) A decoherence-enhanced resonance peak develops around $\varepsilon_\text{C}-\varepsilon_\text{L}$=\unit[-0.3]{meV}, i.e. at the triple crossing (1,1,0)--(0,2,0)--(0,1,1), by the lifting of spin blockade. On the contrary, the dark superexchange state is washed out by decoherence and hence current flows, as shown in the inset.
}
\end{figure}

{\it Spin decoherence.} Dark states are sensitive to decoherence and dephasing~\cite{cpt,darkbell}. When the dark superposition \eqref{darkst} loses its coherence, the states (0,1,1) which are coupled to the drain are populated. Thus the system becomes open to transport  hence leading to a finite current. As investigated in Ref.~\onlinecite{cpt}, measuring the current at the dark state condition provides an estimation of the decoherence rates. Let us consider phenomenological rates accounting for spin relaxation and decoherence, $T_1^{-1}$ and $T_2^{-1}$~\cite{blum}. As shown in Fig.~\ref{curr-rel}, these not only destroy the dark state but also reduce the height of the superexchange resonance line. Both features in fact strongly rely on the coherence in the system. 

On the other hand, finite spin lifetime lifts spin blockade by mixing singlet and triple states. Thus, at the crossing of the states (1,1,0), (0,2,0) and (0,1,1), triplet states --which otherwise contribute to transport via superexchange only-- decay into the resonantly transmitting singlets. As a result, an additional resonance peak appears which becomes sharper with increasing decoherence rates, see Fig.~\ref{curr-rel}(b), a signature of lifetime enhanced coherent transport.

{\it Conclusions.} We propose a triple quantum dot setup within today's experimental reach~\cite{maria,qp12,braakman,amaha} where superexchange interactions can be detected and manipulated. We have found and analyzed current resonances involving only states with merely indirect coupling. The transport mechanism is described in terms of higher-order superexchange transitions. In the resonance of states with left-right inverted charge distribution, an electron is delocalized between the outermost quantum dots with the centre dot being only virtually occupied.
We predict a dark superposition of spin singlets formed at a point of destructive interference of virtual transitions. It is manifested in a total current suppression (the superexchange blockade). 
We emphasize the relevance of spin correlations for systems with more than one electron. 
In particular, resonances which are suppressed by Pauli spin blockade show lifetime-enhanced coherent transport in the presence of decoherence. 
We identify another configuration where left-right symmetric superpositions of spin states are formed which pave the way for CTAP of Bell states.

{\it Acknowledgements.} We acknowledge discussions with G. Granger and A. S. Sachrajda, and financial support from the Spanish MICINN Juan de la  Cierva program (RS) and MAT2011-24331. FGM thanks the hospitality of T. Brandes and the Technische Universt\"at Berlin, where part of this work was accomplished.


\begin{thebibliography}{30}
\bibitem{pauling}
L. Pauling, J. Am. Chem. Soc. {\bf 53}, 1367 (1931).
\bibitem{gray}
H. B. Gray and J. R. Winkler, Proc. Natl. Acad. Sci. USA {\bf 102}, 3534 (2005).
\bibitem{ratner}
M. A. Ratner, J. Phys. Chem. {\bf 94}, 4877 (1990).
\bibitem{hu}
Y. Hu and S. Mukamel, Chem. Phys. Lett. {\bf 160}, 410 (1989).
\bibitem{jortner}
J. Jortner, M. Bixon, T. Langenbacher and M.E. Michel-Beyerle, Proc. Natl. Acad. Sci. USA {\bf 95}, 12759 (1998).
\bibitem{giese}
B. Giese, J. Amaudrut, A.-K. K\"ohler, M. Spormann and S. Wessely, Nature {\bf 412}, 318 (2001).
\bibitem{zener}
C. Zener, Phys. Rev. {\bf 82}, 403 (1951).
\bibitem{anderson1959}
P. W. Anderson, Phys. Rev. {\bf 115}, 2 (1959).
\bibitem{anderson1961}
P. W. Anderson, Phys. Rev. {\bf 124}, 41 (1961).
\bibitem{kondo}
J. Kondo, Prog. Theor. Phys. {\bf 32}, 37 (1964).
\bibitem{andersonCu}
P. W. Anderson, Mat. Res. Bull. {\bf 8}, 153 (1973).
\bibitem{moessner}
R. Moessner and S. L. Sondhi, Phys. Rev. Lett. {\bf 86} 1881 (2001).
\bibitem{barthelemy}
P. Barthelemy and L. M. K. Vandersypen, Ann. Phys. (Berlin) {\bf 525}, 808 (2013).
\bibitem{smirnov}
A. Yu. Smirnov, S. Savel'ev, L. G. Mourokh and F. Nori, Europhys. Lett. {\bf 80}, 67008 (2007).
\bibitem{manousakis}
E. Manousakis, J. Low. Temp. Phys. {\bf 126}, 1501 (2002).
\bibitem{byrnes}
T. Byrnes, N. Y. Kim, K. Kusudo and Y. Yamamoto, Phys. Rev. B {\bf 78}, 075320 (2008).
\bibitem{seo}
M. Seo, H. K. Choi, S.-Y. Lee, N. Kim, Y. Chung, H.-S. Sim, V. Umansky and D. Mahalu, Phys. Rev. Lett. {\bf 110}, 046803 (2013).
\bibitem{gordon}
D. Goldhaber-Gordon, H. Shtrikman, D. Mahalu, D. Abusch-Magder, U. Meirav and M. A. Kastner, Nature {\bf 391}, 156 (1998).
\bibitem{ono}
K. Ono, D.G. Austing, Y. Tokura, S. Tarucha, Science {\bf 297}, 1313 (2002).
\bibitem{gaudreau}
L. Gaudreau, S. A. Studenikin, A. S. Sachrajda, P. Zawadzki, A. Kam, J. Lapointe, M. Korkusinski and P. Hawrylak, Phys. Rev. Lett. {\bf 97}, 036807 (2006).
\bibitem{schroer}
D. Schr\"oer, A. D. Greentree, L. Gaudreau, K. Eberl, L. C. L. Hollenberg, J. P. Kotthaus and S. Ludwig, Phys. Rev. B {\bf 76}, 075306 (2007).
\bibitem{ghislain}
G. Granger, L. Gaudreau, A. Kam, M. Pioro-Ladri\`ere, S. A. Studenikin, Z. R. Wasilewski, P. Zawadzki and A. S. Sachrajda, Phys. Rev. B {\bf 82}, 075304 (2010).
\bibitem{amaha-spin}
S. Amaha, W. Izumida, T. Hatano, S. Teraoka, S. Tarucha, J. A. Gupta and D. G. Austing, Phys. Rev. Lett. {\bf 110}, 016803 (2013).
\bibitem{gaudreau-qubit}
L. Gaudreau, G. Granger, A. Kam, G. C. Aers, S. A. Studenikin, P. Zawadzki, M. Pioro-Ladri\`ere, Z. R. Wasliewszi and A. S. Sachrajda, Nat. Phys. {\bf 8}, 54 (2012).
\bibitem{studenikin}
S. A. Studenikin, G. C. Aers, G. Granger, L. Gaudreau, A. Kam, P. Zawadzki, Z. R. Wasilewski and A. S. Sachrajda, Phys. Rev. Lett. {\bf 108}, 226802 (2012).
\bibitem{laird}
E. A. Laird, J. M. Taylor, D. P. DiVincenzo, C. M. Marcus, M. P. Hanson, and A. C. Gossard, Phys. Rev. B {\bf 82}, 075403 (2010).
\bibitem{maria}
M. Busl, G. Granger, L. Gaudreau, R. S\'anchez, A. Kam, M. Pioro-Ladri\`ere, S. A. Studenikin, P. Zawadzki, Z. R. Wasilewski, A. S. Sachrajda and G. Platero, Nat. Nanotechnol. {\bf 8}, 261 (2013).
\bibitem{qp12}
R. S\'anchez, G. Granger, L. Gaudreau, A. Kam, M. Pioro-Ladri\`ere, S. A. Studenikin, P. Zawadzki, A. S. Sachrajda and G. Platero, arXiv:1312.5060.
\bibitem{braakman}
F. R. Braakman, P. Barthelemy, C. Reichl, W. Wegscheider and L. M. K. Vandersypen, Nat. Nanotechnol. {\bf 8}, 432 (2013).
\bibitem{reilly}
D. J. Reilly, Nat. Nanotechnol. {\bf 8}, 395 (2013).
\bibitem{saraga}
D. S. Saraga and D. Loss, Phys. Rev. Lett. {\bf 90}, 166803 (2003).
\bibitem{mariaSB}
M. Busl, R. S\'anchez and G. Platero, Phys. Rev. B {\bf 81}, 121306 (2010).
\bibitem{amaha}
S. Amaha, T. Hatano, H. Tamura, S. Teraoka, T. Kubo, Y. Tokura, D. G. Austing and S. Tarucha, Phys. Rev. B {\bf 85}, 081301 (2012).
\bibitem{fabian}
J. Fabian and U. Hohenester, Phys. Rev. B {\bf 72}, 201304 (2005).
\bibitem{jan}
J. Huneke, G. Platero and S. Kohler, Phys. Rev. Lett. {\bf 110}, 036802 (2013).
\bibitem{averin}
D. V. Averin and Yu. V. Nazarov, Phys. Phys. Lett. {\bf 65}, 2446 (1990).
\bibitem{alfredo}
A. Levy Yeyati and M. B\"uttiker, Phys. Rev. B {\bf 62}, 7307 (2000).
\bibitem{cpt}
B. Michaelis, C. Emary and C. W. J. Beenakker, Europhys. Lett. {\bf 73}, 677 (2006).
\bibitem{brandes}
T. Brandes and F. Renzoni, Phys. Rev. Lett. {\bf 85}, 4148 (2000).
\bibitem{darkbell}
R. S\'anchez and G. Platero, Phys. Rev. B {\bf 87}, 081305 (2013).
\bibitem{blum}
K. Blum, {\it Density Matrix Theory and Applications} (Plenum, New York, 1996).

\end{thebibliography}
\end{document}